\documentclass{article}
\usepackage{spconf}
\usepackage{amsmath,graphicx,amssymb,bm,units,subfigure,acronym,color,enumerate,balance,graphics,graphicx,psfrag,blindtext,ctable,epstopdf}
\usepackage{caption, balance}
\usepackage[english]{babel}
\usepackage[sort]{cite}

\title{Multi-scale aggregation of phase information for reducing computational cost of CNN based DOA estimation}
\name{Soumitro Chakrabarty and Emanu\"{e}l A. P. Habets}
\address{International Audio Laboratories Erlangen\sthanks{\,A joint institution of the Friedrich-Alexander-University Erlangen-N\"urnberg (FAU) and Fraunhofer Institute for Integrated Circuits (IIS).},  
	Am Wolfsmantel 33, 91058 Erlangen, Germany\\
	\{soumitro.chakrabarty, emanuel.habets\}@audiolabs-erlangen.de}
\begin{document}
\ninept
\maketitle
\begin{abstract}
In a recent work on direction-of-arrival (DOA) estimation of multiple speakers with convolutional neural networks (CNNs), the phase component of short-time Fourier transform (STFT) coefficients of the microphone signal is given as input and small filters are used to learn the phase relations between neighboring microphones. Due to this chosen filter size, $M-1$ convolution layers are required to achieve the best performance for a microphone array with M microphones. For arrays with large number of microphones, this requirement leads to a high computational cost making the method practically infeasible. In this work, we propose to use systematic dilations of the convolution filters in each of the convolution layers of the previously proposed CNN for expansion of the receptive field of the filters to reduce the computational cost of the method. Different strategies for expansion of the receptive field of the filters for a specific microphone array are explored. With experimental analysis of the different strategies, it is shown that an aggressive expansion strategy results in a considerable reduction in computational cost while a relatively gradual expansion of the receptive field exhibits the best DOA estimation performance along with reduction in the computational cost.
\end{abstract}
\section{Introduction}
\label{sec:intro}

Many applications such as hands-free communication, teleconferencing, robot audition and distant speech recognition require information on the location of sound sources in the acoustic environment. The relative direction of a sound source with respect to a microphone array is generally given in terms of the direction of arrival (DOA) of the sound wave originating from the source position.  In most practical scenarios, this information is not available and the DOA of the sound source need to be estimated.  

Compared to signal processing based approaches to this task, supervised learning approaches have the advantage that they can be adapted to different acoustic conditions via training. With the recent success of deep neural network based supervised learning methods for different signal processing related tasks, they have also become an attractive solution for DOA estimation \cite{Ma2017, Vesperini2016, Takeda2016, Xiao2015, Takeda2016a, Chakrabarty2017a, Chakrabarty2017b, Adavanne2018, Perotin2018, Chakrabarty2018}. 

Existing approaches mainly vary in terms of the input features that are utilized for the task of DOA estimation. Most of the earlier methods \cite{Ma2017, Vesperini2016, Takeda2016, Xiao2015, Takeda2016a} involved a feature extraction step, where features similar to those used in classical signal processing based approaches, were given as an input to a deep neural network to learn the mapping from the features to the DOA of the sound sources. 

The current authors proposed a convolutional neural network (CNN) based supervised learning method for broadband DOA estimation where the phase component of short-time Fourier transform (STFT) coefficients of the input signal were directly provided as input to the neural network \cite{Chakrabarty2017a}. Following this, other methods were also proposed that use the time-frequency representation of the recorded microphone signals as input to the neural network and utilized convolution layers in the early part of the proposed architecture to learn the required features from the input for the DOA estimation task \cite{Chakrabarty2017b,Chakrabarty2018, Adavanne2018}.  

In \cite{Chakrabarty2018} a CNN based DOA estimation framework for the task of estimating the DOAs of multiple simultaneously active sound sources was proposed by the current authors. In that work, an experimental analysis of the number of required convolution layers for the best performance of the proposed framework was presented and it was found that for a uniform linear array (ULA) with $M$ microphones, $M-1$ convolution layers are required due to the choice of small filters that learn the phase relations from neighboring microphones. Such a requirement limits the applicability for microphone arrays with large number of microphones, since it leads to a high computational cost due to the large number of convolution layers required to achieve a good DOA estimation performance. Also, a large number of convolution layers leads to a very deep network which can lead to the vanishing gradient problem \cite{Schmidhuber2015}. In \cite{Chakrabarty2018}, it was also shown that by simply reducing the number of convolution layers, the proposed method suffered from considerable degradation in performance.   
 
In this paper, we propose to incorporate systematic dilation of the filters in each of the convolution layers to reduce the requirement of $M-1$ convolution layers in the proposed framework in \cite{Chakrabarty2018}. The idea of using systematic dilated convolutions for multi-scale aggregation of contextual information in the feature space was first introduced in \cite{Yu2016} for dense prediction tasks in computer vision. In \cite{Yu2016}, it was demonstrated that dilated convolutions allow for the exponential expansion of the receptive field of the filters without loss of resolution. In audio-related tasks, dilated convolutions are mainly used in generative models for audio synthesis \cite{Oord2016}. In this work, we utilize systematic dilated convolutions for aggressive expansion of the receptive field of the convolution filters such that phase information from all the microphones can be aggregated in fewer than $M-1$ convolution layers. Different strategies for the incorporation of dilated convolutions within the convolution layers are investigated for a specific microphone array and the reduction in computational cost that can be achieved is presented. We also investigate the use of larger filters as a possible solution to this problem. Finally, the DOA estimation performance of the proposed modifications to the CNN presented in \cite{Chakrabarty2018} is compared to the original architecture.  

\section{DOA estimation with CNNs}
\label{sec:doa_cnn}

In this first part, a brief overview of the previously proposed CNN based framework \cite{Chakrabarty2017b} for estimating the DOAs of multiple simultaneously active speakers is presented. 

DOA estimation in the mentioned work is performed for signal blocks that consist of multiple time frames of the Short-time Fourier Transform (STFT) representation of the observed signals, where the block length can be chosen based on the application scenario. 

The problem is formulated as a multi-label multi-class classification problem, where the complete DOA range is discretized to form the DOA class labels. The CNN learns to assign multiple DOA class labels to the input corresponding to each time frame using a large amount of training data. In the test phase, given the input feature representation corresponding to a single STFT time frame, the first task is to estimate the posterior probability of each DOA class. The assignment of each DOA class label is treated as a separate binary classification problem, assuming an independent source location model. Following this, depending on the chosen block length, the frame-level probabilities are averaged over all the time frames in the block. Finally, considering $L$ sources, the DOA estimates are given by selecting the $L$ DOA classes with the highest probabilities. 

The input feature representation in the framework is the \emph{phase map}, that was first introduced in \cite{Chakrabarty2017a}. The phase map, for the $n$-th time frame is formed by arranging the phase of the STFT coefficients for each time-frequency bin $(n,k)$ and each microphone $m$ into a matrix of size $K \times M$, where $K = N_{f}/2 +1$ is the total number of frequency bins, upto the Nyquist frequency, at each time frame and $M$ is the total number of microphones in the array.

Given the phase map as the input, the CNN generates the posterior probability for each of the DOA classes. In the convolution layers of the CNN, small filters of size $1 \times 2$ are applied to learn the phase relations between neighboring microphones at each frequency sub-band separately. These learned features for each sub-band are then aggregated by two fully connected layers leading to the output for the classification task. 

An important design aspect of the previously proposed architecture is the number of convolution layers. In \cite{Chakrabarty2018}, it was experimentally demonstrated that $M-1$ convolution layers are required to obtain the best DOA estimation performance for a microphone array with $M$ microphones. This result can be attributed to the fact that by using small filters of size $1 \times 2$, with each subsequent convolution layer after the first one, for each sub-band, the phase correlation information from different microphone pairs are aggregated by the growing receptive field of the filters, and to learn from the correlation between all microphone pairs, $M-1$ convolution layers are required to incorporate this information into the learned features. 

One of the main drawbacks of this requirement is that for arrays with large number of microphones, the number of required convolution layers becomes high leading to a large computational requirement, which can become practically infeasible. Therefore, in this work we investigate possible modifications to this previously proposed architecture to reduce the computational requirement while still obtaining good DOA estimation performance. 

\section{Receptive field expansion with Dilated Convolutions}
\label{sec:pagestyle}

A main reason for the requirement of $M-1$ convolution layers in the architecture proposed in \cite{Chakrabarty2018} is the gradual aggregation of information in the feature space by the slowly growing receptive field of the small filters used in the framework. A possible solution for this problem is to use larger filters, however this can lead to an increase in the computational cost as well as the number of trainable parameters. In this work, we propose to incorporate systematic dilation of the convolution filters/kernels \cite{Yu2016} to expand the receptive field of the filters with each convolution layer. 

\subsection{Dilated Convolutions}
\label{sec:dil}

In the generic discrete convolution operation in CNNs, filters are applied to learn from adjacent local parts of the input. This is also known as contiguous convolutions. In dilated convolutions, based on the dilation factor, $R$, the applied filters are able to learn from distant elements in the input space while having the same number of filter elements. It should be noted that in the case of 2D convolutions, different dilation factors can be applied along the different dimensions of the filter. 

\begin{figure}[t]
	\captionsetup{width=.48\textwidth}
	\centering
	\subfigure[Contiguous convolution]{\label{subfig:cont}\includegraphics[width=0.2\textwidth]{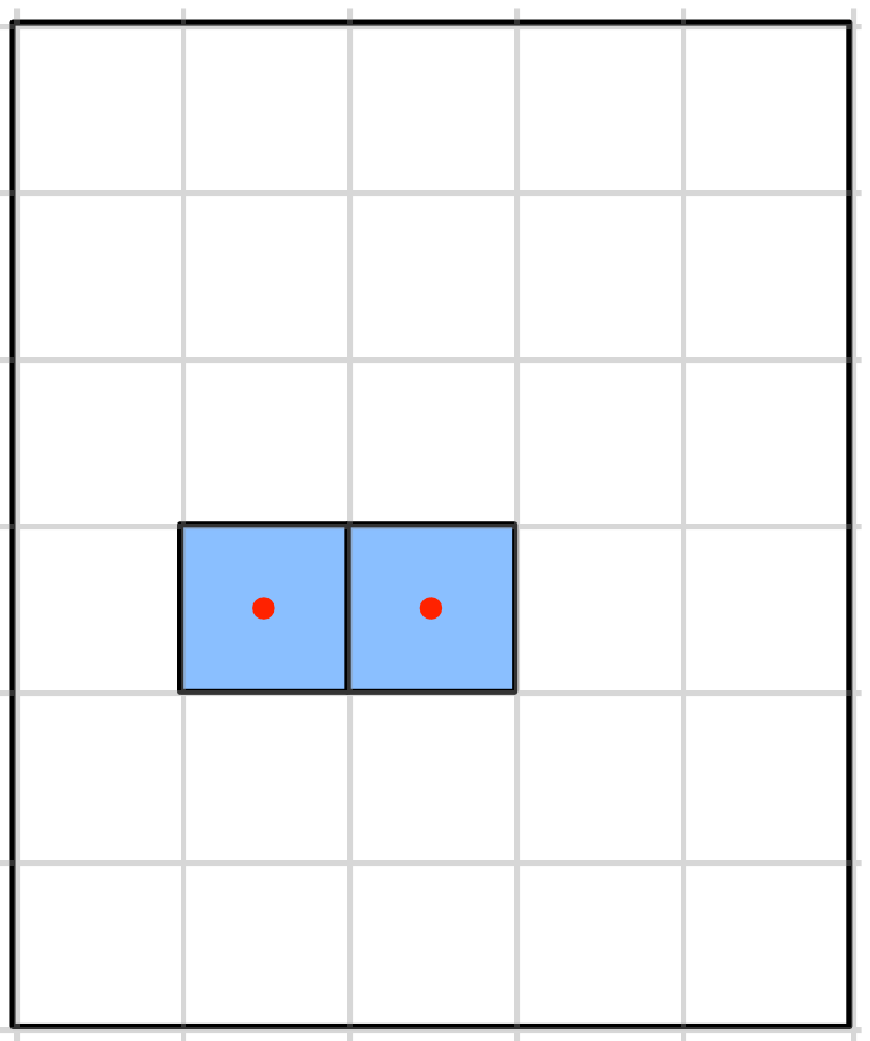}}
	\hspace{3 em}
	\subfigure[Dilated convolution with dilation factor of $R = (1,2)$]{\label{subfig:dil}\includegraphics[width=0.2\textwidth]{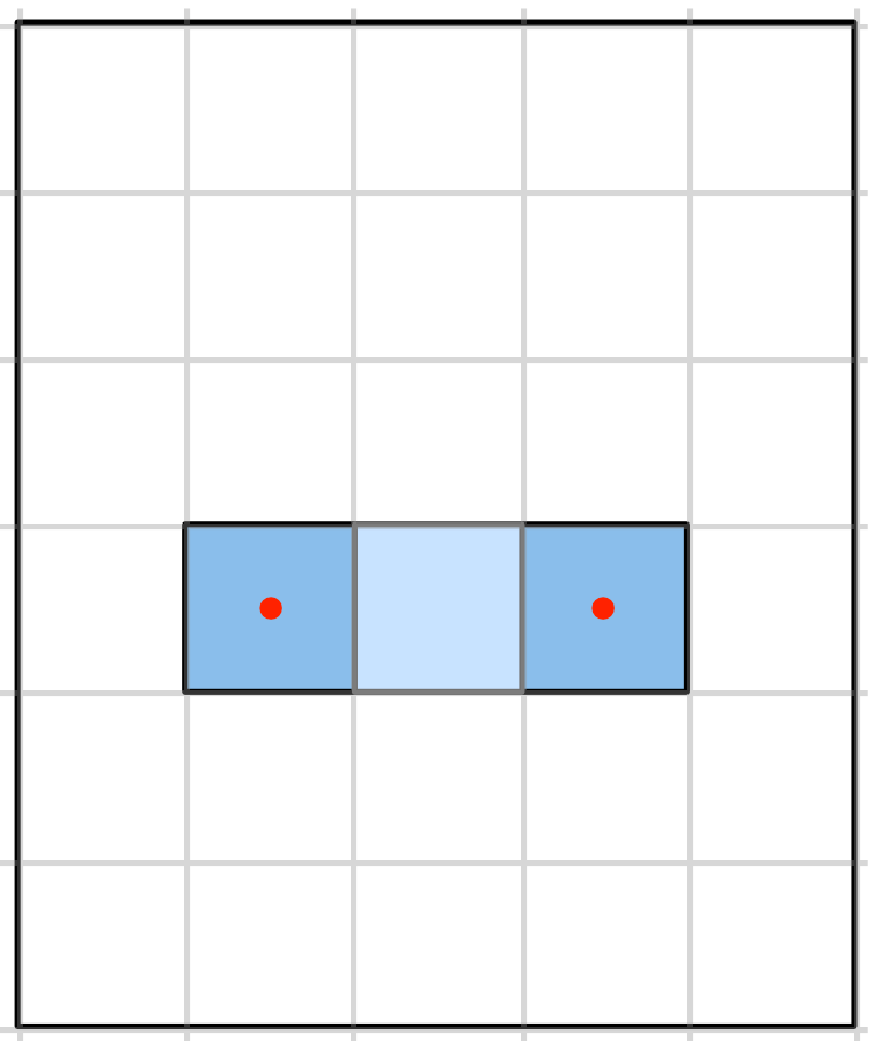}}
	%
	\caption{Illustrative example of contiguous and dilated convolution.}
	\label{fig:Dil}
\end{figure}

An illustrative example, in context of the filters used in our proposed architecture in \cite{Chakrabarty2018}, of the difference between contiguous convolution and dilated convolution is shown in Fig.~\ref{fig:Dil}. In Fig.~\ref{subfig:cont}, the basic discrete convolution is shown where the filter is applied to two adjacent elements of the input matrix, while Fig.~\ref{subfig:dil} shows a dilated convolution operation with the same number of filter elements where a dilation factor of 2 is applied along the column dimension of the filter. The dilation factor mainly determines the size of the gap between the filter elements along a certain dimension. It should be noted that a dilation factor of 1 leads to the basic discrete convolution operation.      

One of the main advantages of dilated convolutions is that with systematic inclusion of dilations an exponential growth in the receptive field of the filters with each convolution layer can be achieved, while the number of parameters grow linearly. For further details regarding dilated convolutions, we refer the readers to \cite{Yu2016}.        

\begin{table*}[t]
	\small
	\centering
	\setlength{\tabcolsep}{1.7mm}
	\vspace{1 em}
	\begin{tabular}{l  c c c c c c c c}
		\toprule
		Architectures & & Conv. Layers& & FLOPs ($\times 10^{6}$)&  & FLOPs (w.r.t \cite{Chakrabarty2018})& & Tr. Parameters ($\times 10^{6}$)\\
		\midrule 
		\cite{Chakrabarty2018} & &7& &$53.14$ & &1& & $8.75$  \\	
		F2342   			   & &4& &$35.24$ & &0.66& & $8.84$    \\
		D1123   			   & &4& &$32.08$ & &0.60& & $8.73$    \\
		D133    			   & &3& &$19.45$ & &0.36& & $8.72$    \\
		\bottomrule
	\end{tabular}
	\caption{Number of convolution layers, number of FLOPs and number of trainable parameters for the different compared architectures.}
	\label{tab:run}\vspace{0.1em}
\end{table*}
\subsection{Design Considerations}
\label{ssec:design}

The main aim in this work is to utilize the aggressive expansion of receptive field afforded by using dilated convolutions to reduce the requirement of $M-1$ convolution layers within our previously proposed architecture. Keeping the number of elements in the filters same, the main design choice is the dilation factor for each convolution layer. 

One of the first things to note is that in the first convolution layer the dilation factor should be 1. This is required to avoid any loss of resolution in the feature space for the learned filters, since a dilation factor of greater than 1 restricts the filters from learning from phase relations between the neighboring microphones. In preliminary experiments, significant reduction in performance was observed when using a dilation factor of greater than 1 in the first layer.

The choice of dilation factor for the subsequent layers depends on the desired reduction in the number of convolution layers. With a specific choice of number of convolution layers, the dilation factors of all the convolution layers should sum up to $M-1$ such that the microphone dimension of the output feature map after the last convolution layers is 1. This keeps the number of trainable parameters manageable while also having the filters cover the whole microphone dimension of the input feature space.    
\begin{table}[t]
	\footnotesize
	\centering
	\setlength{\tabcolsep}{3.2 mm}
	\vspace{1 em}
	\begin{tabular}{l | l  c}
		\toprule
		Signal & Speech signals from LIBRI \\
		
		Room size  & Room 1: ($4 \times 7$) m , Room 2: ($9 \times 7$) m \\
		
		Array positions in room & 3 arbitrary positions in each room \\
		
		Source-array distance& 1.3 m  for Room 1, 2.1 m for Room 2\\		  
		
		RT$_{60}$ & Room 1: 0.38 s , Room 2: 0.52 s \\	
		
		SNR           & 20 dB and 30 dB  \\
		\bottomrule
	\end{tabular}
	\caption{Configuration for generating test data for the experiments. All rooms are 3 m high.} 
	\label{tab:Test}\vspace{0.1em}
\end{table}
\section{Experimental Validation}
\label{sec:exp}

The design modifications proposed in the previous section are experimentally validated in this section with a specific microphone array for different acoustic conditions. The performance is also compared to the basic architecture from \cite{Chakrabarty2018} to investigate the difference in performance due to the propsoed modifications.

\subsection{Experimental Setup}
\label{ssec:setup}
For the experimental investigation, we consider a ULA with $M = 8$ microphones with inter-microphone distance of 2 cm, and the input signals are transformed to the STFT domain using a DFT length of $N_{f} = 512$, with $50\%$ overlap, resulting in $K = 257$. The sampling frequency of the signals is $F_{s} = 16$ kHz. To form the classes, we discretize the whole DOA range of a ULA with a $5^{\circ}$ resolution to get $I = 37$ DOA classes, for both training and testing. All the presented objective evaluations are for the two speakers scenario.

The speech signals used for evaluation are taken from the LibriSpeech corpus \cite{Panayotov2015}. Since the angular space is discretized with a 5$^{\circ}$ resolution it was ensured that the angular distance between the two simultaneously active speakers is at least 10$^{\circ}$. During test, for a specific source-array setup in a room, a two speaker mixture is considered for all possible angular combination. This was done to avoid the influence of signal variation on the difference in performance for different acoustic conditions. 

Since the speech utterances can have different lengths of silence at the beginning, the central 0.8 s segment of the mixtures was selected for evaluation. Considering an STFT window length of 32 ms with 50$\%$ overlap, this resulted in a signal block of $N= 50$ time frames over which the frame-level posterior probabilities are averaged for the final DOA estimation. 

For evaluation, two different objective measures were used: Mean Absolute Error (MAE) and localization accuracy (Acc.) \cite{Chakrabarty2018}.

\subsection{Compared Architectures}
\label{ssec:comp_arch}

To experimentally investigate the difference in performance due to different design choices regarding the convolution filters, we compare the performances of the following architectures:

\begin{itemize}
	\item CNN with contiguous convolutions proposed in \cite{Chakrabarty2018}, with different number of convolution layers ranging from 2 to $M-1 =7$. This is referred as the baseline architecture.
	\item CNN with contiguous convolutions with larger filters after the first convolution layer. We reduce the number of convolution layers to be 4, with filters of size 2 for the first layer, followed by filters of size 3,4 and 2. This architecture is referred as F2342.
	\item CNN with dilated convolutions and 4 convolution layers. The dilation factors starting from the first convolution layer are 1,1,2 and 3.  This architecture is referred as D1123.
	\item CNN with dilated convolutions and 3 convolution layers. The dilation factors starting from the first convolution layer are 1,3 and 3.  This architecture is referred as D133.
\end{itemize}

In total, 9 different CNNs were trained for this experiment. The number of fully connected layers, as well as the activation functions were same for all the networks. All the networks were trained with the same amount of data. Multi-condition training with simulated training data for diverse acoustic conditions, same as in \cite{Chakrabarty2018}, was performed for robust performance in different acoustic scenarios. 


\begin{figure*}[t]
	\captionsetup{width=.99\textwidth}
	\centering
	\subfigure[MAE]{\label{subfig:maeexp3}\includegraphics[width=0.44\textwidth]{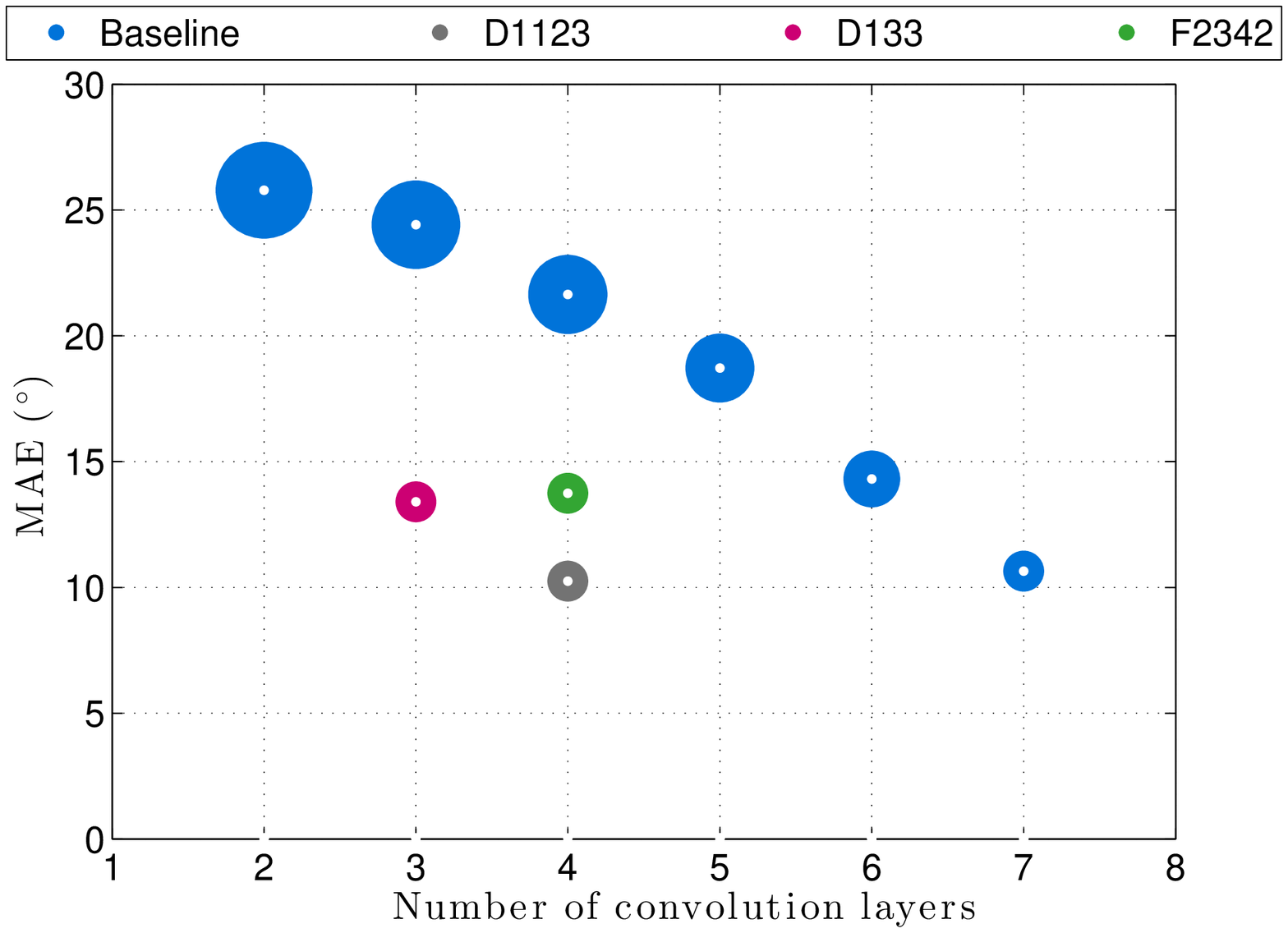}}
	\hspace{3 em}
	\subfigure[Accuracy]{\label{subfig:accexp3}\includegraphics[width=0.447\textwidth]{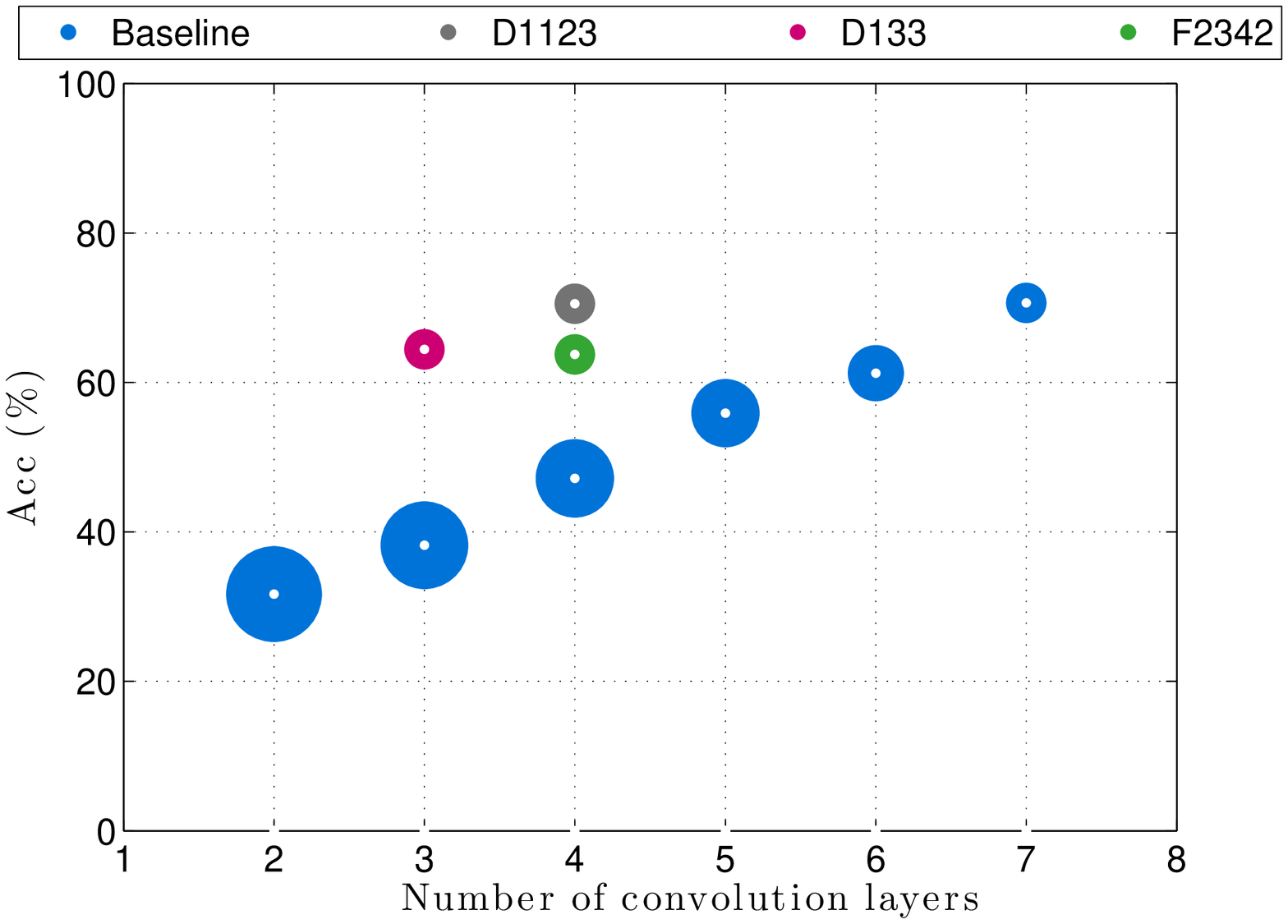}}
	%
	\caption{DOA estimation performance of the different compared architectures.}
	\label{fig:Exp}
\end{figure*} 

\subsection{Computation and Memory Requirement}
\label{ssec:flops}

In this section, we look at the computation and memory requirements for the different architectures introduced in the previous section. The computational requirement is presented in terms of the number of floating point operations (FLOPs) required for a single forward pass of the input phase map through the network. For each convolution layer, the FLOPs are computed as the product of three dimensions (height, width , depth) of the input map and the dimensions of the filters. For the fully connected layers, the number of FLOPS is given as a product of the input and the output vector lengths. The memory requirement is given in terms of the total number of trainable parameters (including bias terms) in each architecture. 

The number of convolution layers, and the computation and memory requirements for the different compared architectures is shown in Table~\ref{tab:run}. For the baseline architecture, the computation and memory requirement for only the network with $M-1 = 7$ convolution layers is shown.   

It can be seen in Table~\ref{tab:run}, that the memory requirement for the different architectures are quite similar. The network with larger filters (F2342) has the highest memory requirement due to the application of larger filters. For the networks with dilated convolutions, the number of trainable parameters is slightly lower than the baseline architecture, due to the reduced number of convolution layers.  

Though the memory requirement for the different architectures is similar, the number FLOPs for the networks with dilated convolutions (D1123, D133) and larger filters (F2342) are much lower than the previously proposed architecture (Table~\ref{tab:run}). For the network D133, where higher dilation factor is applied in the early convolution layers for a more aggressive expansion of the receptive field of the filters, the number of FLOPs is almost a third of that of the baseline architecture. For the network with dilated convolutions with a more gradual expansion of the receptive field (D1123), the number of FLOPs is 60$\%$ of the baseline architecture. It can also be seen that by using larger filters, the number of FLOPs can be reduced by $35\%$ for the microphone array setup considered here.  

Therefore, we see that by introducing the proposed modifications to the network, a considerable reduction in computational requirement can be achieved while keeping the number of trainable parameters similar to the baseline architecture. However, the influence of these modifications on the DOA estimation performance of the network needs to be evaluated, which is presented in the next section. 

\subsection{Results}
\label{ssec:results}

The DOA estimation performance of the 9 different trained networks was evaluated with simulated data. The acoustic conditions used for test are	 shown in Table.~\ref{tab:Test}. 

The performance of the different compared architectures, in terms of both MAE and localization accuracy, is shown in Fig. \ref{fig:Exp}. In the figures, the center of the circle markers correspond to the value of the objective measure and the area of the markers denote the number of trainable parameters for that specific architecture. Due to space constraints, we present results averaged over the different acoustic conditions given in Table~\ref{tab:Test}. 

For the baseline architecture, it can be seen that by simply reducing the number of convolution layers, there is considerable degradation in performance (blue in Fig.~\ref{fig:Exp}). By using larger filters with 4 convolution layers (F2342, green in Fig.~\ref{fig:Exp}), an improvement of $8^\circ$ in terms of MAE and $16\%$ in terms of accuracy can be seen compared to the baseline architecture with 4 convolution layers. Using dilated convolutions and an aggressive receptive field expansion strategy (D133, pink in Fig.~\ref{fig:Exp}), slightly better performance than F2342 is achieved with only 3 convolution layers but compared to the baselineline architecture with 7 layers, there is a loss of $7\%$ in terms of accuracy and the MAE increases by $4^\circ$. The best performance is achieved by the D1123 network (gray in Fig.~\ref{fig:Exp}), which consists of 4 convolution layers and uses a gradual expansion of the receptive field.  

From the results, we see that by using systematic dilations with a gradual expansion strategy (D1123), a performance similar to the baseline network with $M-1$ convolution layers is achieved for the considered scenario, with a reduction of $40\%$ in FLOPs. With an aggressive expansion strategy for the receptive field of the filters (D133), the computation requirement can be reduced by $64\%$, however there is a 7$\%$ loss in accuracy and $4^\circ$ increase in MAE compared to the baseline network with $M-1$ layers. 

\section{Conclusion}

We proposed the incorporation of systematic dilation of the convolution filters to reduce the computational requirement of our previously proposed network for DOA estimation by expanding the receptive field of the filters in the convolution layers. It was shown that by utilizing systematic dilation the computational cost can reduced while keeping the memory requirement similar to the original network. Through experimental analysis with a specific microphone array in different acoustic conditions, it was found that though an aggressive expansion of the receptive field of the filters leads to almost $65\%$ reduction in the computational cost, it suffers from degradation in performance compared to the original architecture. With a more gradual expansion of the receptive field, similar performance to the original network with $M-1$ layers can be achieved while reducing the computational cost by 40$\%$. 
       
\balance
\bibliographystyle{IEEEbib}
\bibliography{sapref_icassp19}

\end{document}